\documentclass[english,aps,manuscript]{revtex4}
\usepackage[LGR,T1]{fontenc}
\usepackage[latin9]{inputenc}
\usepackage{xcolor}
\usepackage{pdfcolmk}
\usepackage{units}
\usepackage{textcomp}
\usepackage{amsmath}
\usepackage{amssymb}
\usepackage{graphicx}
\usepackage{esint}
\PassOptionsToPackage{normalem}{ulem}
\usepackage{ulem}

\makeatletter

\DeclareRobustCommand{\greektext}{%
  \fontencoding{LGR}\selectfont\def\encodingdefault{LGR}}
\DeclareRobustCommand{\textgreek}[1]{\leavevmode{\greektext #1}}
\DeclareFontEncoding{LGR}{}{}
\DeclareTextSymbol{\~}{LGR}{126}
\newcommand{\lyxmathsym}[1]{\ifmmode\begingroup\def\b@ld{bold}
  \text{\ifx\math@version\b@ld\bfseries\fi#1}\endgroup\else#1\fi}

\providecolor{lyxadded}{rgb}{0,0,1}
\providecolor{lyxdeleted}{rgb}{1,0,0}
\@ifundefined{textcolor}{}
{%
 \definecolor{BLACK}{gray}{0}
 \definecolor{WHITE}{gray}{1}
 \definecolor{RED}{rgb}{1,0,0}
 \definecolor{GREEN}{rgb}{0,1,0}
 \definecolor{BLUE}{rgb}{0,0,1}
 \definecolor{CYAN}{cmyk}{1,0,0,0}
 \definecolor{MAGENTA}{cmyk}{0,1,0,0}
 \definecolor{YELLOW}{cmyk}{0,0,1,0}
}

\makeatother

\usepackage{babel}
\begin{document}

\title{Particle Dynamics and Rapid Trapping in Electro-Osmotic Flow Around
a Sharp Microchannel Corner}

\author{Matan Zehavi and Gilad Yossifon}

\affiliation{Faculty of Mechanical Engineering, Micro and Nanofluidics Laboratory,
Technion\textendash{}Israel Institute of Technology, Technion City
32000, Israel}
\begin{abstract}
We study here the curious particle dynamics resulting from electro-osmotic
flow around a microchannel junction corner whose dielectric walls
are weakly polarizable. The hydrodynamic velocity field is obtained
via superposition of a linear irrotational term associated with the
equilibrium zeta potentials of both the microchannel and particle
surfaces and the non-linear induced-charge electro-osmotic flow which
originates from the interaction of the externally applied electric
field on the charge cloud it induces at the solid-liquid interface.
The particle dynamics are analyzed by considering dielectrophoretic
forces via the addition of a mobility term to the flow field in the
limit of Stokes drag law. The former, non-divergence free term is
responsible for migration of particles towards the sharp microchannel
junction corner, where they can potentially accumulate. Experimental
observations of particle trapping for various applied electric fields
and microparticle size are rationalized in terms of the growing relative
importance of the dielectrophoretic force and induced-charge contributions
to the global velocity field with increasing intensity of the externally
applied electric field. 
\end{abstract}

\address{{*}Corresponding author: yossifon@tx.technion.ac.il}

\maketitle

\section{Introduction}

The present contribution focuses on the dynamics of particles and
their accumulation in electro-osmotic flow through micro-channel junctions
which are a basic element of microfluidic systems. This work thus
extends previous studies \cite{takhistov_electrokinetic_2003,thamida_nonlinear_2002,yossifon_electro-osmotic_2006,eckstein_nonlinear_2009}
that focused on describing the interesting hydrodynamic behavior,
wherein beyond a critical level of external-field intensity, vortices
are observed to appear within the flow around (sharp) corners of micro-channel
junctions. The generation of such vortices is potentially useful in
certain applications as a means to enhance and control micro-fluid
mixing \cite{wu_micromixing_2008,chen_vortex_2008,zhao_microfluidic_2007}.
In other situations the appearance of vortices may need to be suppressed
so as to avoid accumulation of suspended particles leading to the
eventual jamming of the device \cite{thamida_nonlinear_2002}. In
linear electro-osmosis, when the equilibrium zeta potential is uniform
and independent of the external field, the resulting flow is irrotational.
However, the small yet finite polarizability of the walls gives rise
to an additional, non-linear, electro-kinetic mechanism termed induced-charge
electro-osmosis (ICEO) \cite{squires_induced-charge_2004}. When an
external field is applied to the system, it sets off within the electrolyte
solution transient Ohmic currents which create a non-uniform charge
cloud near the walls. Thus, a non-uniform distribution of induced
zeta potential is established whose magnitude is proportional to the
external-field intensity. The Helmholtz-Smoluchowski slip velocity
at the fluid-solid interface resulting from the interaction of the
electric field and the induced-charge cloud is non-linear in the applied
field and generates an electro-osmotic flow which is not necessarily
irrotational. 

Thamida \& Chang \cite{thamida_nonlinear_2002} pointed out that the
above mechanism produces opposite polarization of the respective upstream
and down-stream faces of the corner. Furthermore, for a sufficiently
strong external field this ICEO not only dominates the local flow
near the corner but produces downstream macro-scale vortices as well
\cite{yossifon_electro-osmotic_2006}. Thamida \& Chang \cite{thamida_nonlinear_2002}
also observed that an interesting colloid accumulation occurs at the
corner, however no attempt was made to address this particle trapping
mechanism either theoretically or experimentally.

As will be shown herein, the rapid trapping occurs due to a combination
of the short range DEP trapping force and the long-range induced electro-osmotic
flow which feeds the particles. Previous studies have examined particle
trapping and separation based on insulator-based dielectrophoresis
(iDEP) using sharp tips among other geometries \cite{staton_characterization_2010,srivastava_dc_2011,liao_nanoscale_2012}.
However, in contrast to the current study, these works have treated
such structures as insulators and as a result induced electrokinetic
effects, emanating from the finite but small dielectric polarizability
of the wall materials, were overlooked. Nevertheless, efficient trapping
via the combination of a short range DEP force and far-field hydrodynamics
has been shown for a variety of flow generation mechanisms. AC electroosmosis
(ACEO), which uses electrodes embedded within the microchannel, has
attracted a lot of attention as means for vortex generation and particle
trapping through hydrodynamic forces that focus the particles into
the stagnation point of the vortices \cite{ramos_ac_1998,ramos_ac_1999,wu_long-range_2005,hoettges_use_2003}.
Additionally, similar accumulation has recently been demonstrated
\cite{green_dynamical_2013} at the stagnation points of electro-osmotic
vortices of the second kind formed at the interface of a microchannel
and a wide nanoslot.

It is thus the focus of the current study to provide for the first
time a thorough explanation, based on both theory and experiments,
for the rapid particle trapping due to the combined forces of the
DEP and ICEO vortex at the corner of a microchannel. In the following,
we describe the experimental methods in sec. 2, the theoretical model
in sec. 3, the results and discussion in sec. 4 and concluding comments
in sec. 5.

\section{Experimental Methods}

\subsection{Fabrication of the device}

\begin{figure}
\begin{centering}
\includegraphics[width=0.95\columnwidth]{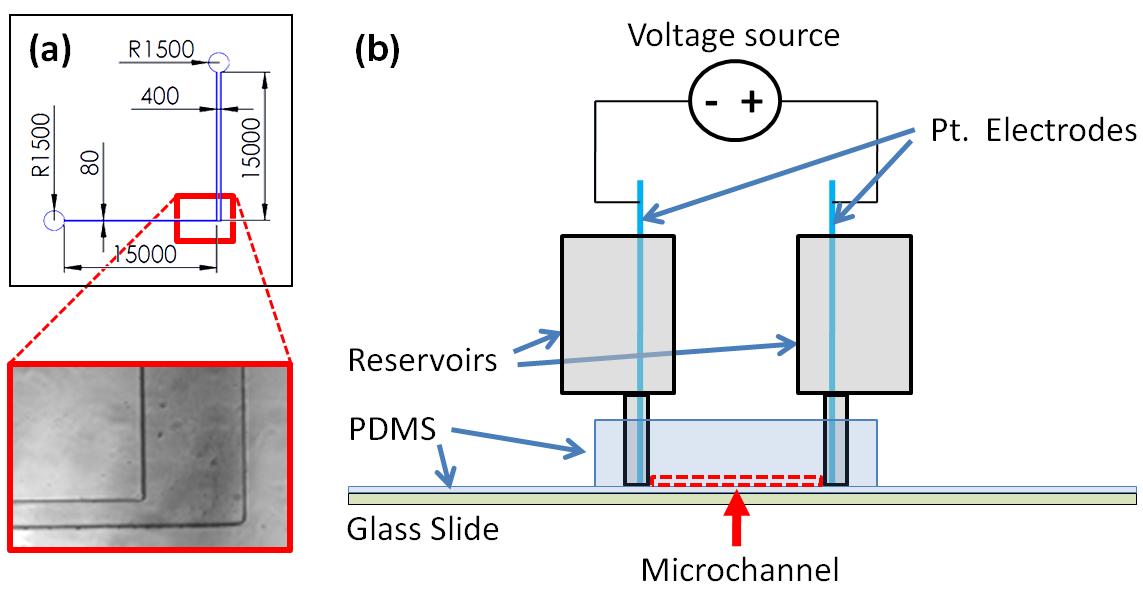}
\par\end{centering}

\caption{ (a) Schematics of the L-junction microchannel device along with a
microscope image of the junction region and (b) the experimental flow
system setup. Dimensions are in \textgreek{m}m.\label{fig: the flow system}}
\end{figure}

A L-junction microchannel connected to two reservoirs at opposite
ends (Fig. \ref{fig: the flow system}(a))
 with a depth (normal to the plane of view) of 120 \textmu{}m was
fabricated from PDMS (Polydimethylsiloxane, Dow-corning Sylgard 184)
using a rapid prototype technique \cite{anderson_fabrication_2000}.
We used a high resolution (3 \textgreek{m}m) chrome mask for the creation
of the master. The radii of curvature of the corner is estimated to
be \textasciitilde{}6 \textmu{}m The channel was then sealed by a
PDMS coated (30 \textmu{}m thick) microscope glass slide, using a
plasma bonding process \cite{haubert_pdms_2006}. Two large (19 mm
diameter) reservoirs were inserted into the PDMS inlets so as to minimize
possible pressure driven flow due to induced pressure head and to
allow the introduction of Platinum wire electrodes (Fig. (\ref{fig: the flow system})b).

\subsection{Experimental setup}

A high voltage DC power supply (Stanford Research Systems PS350) was
connected to the platinum wire electrodes (0.5 mm platinum wire, Sigma-Aldrich).
An electrolyte solution of 10\textsuperscript{-5}{[}M{]} Potassium
Chloride (conductivity $\sigma_{f}=2$ \textgreek{m}S/cm) was seeded
by negatively charged fluorescently tagged tracer particles (Fluoro-Max,
Thermo-Scientific) of various sizes (0.48, 1 and 2 \textgreek{m}m)
and volumetric concentrations of (0.0025\%, 0.01\%, 0.04\% respectively).
The particles were visualized dynamically using a spinning disc confocal
system (Yokugawa CSU-X1) connected to a camera (Andor iXon3) and installed
on an inverted microscope (Nikon TI Eclipse). Prior to the sudden
application of the external field the system was equilibrated to minimize
initial pressure driven flow.

\section{Theoretical model}

\subsection{Electrostatics and hydrodynamics }

\begin{figure}
\begin{centering}
\includegraphics[width=0.5\columnwidth]{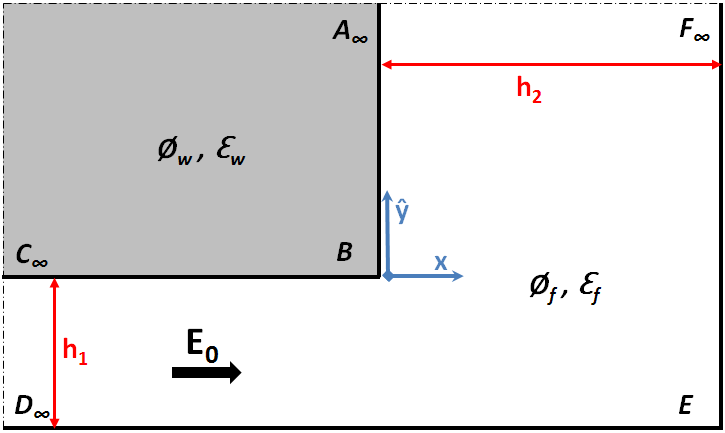}
\par\end{centering}

\caption{\label{fig:The-L-junction-configuration.}The L-junction configuration}
\end{figure}

Here we extend the theoretical analysis of Yossifon et al. \cite{yossifon_electro-osmotic_2006}
for ICEO around a sharp corner to include DEP forces acting on particles.
For brevity, in deriving the latter hydrodynamic contribution to the
particle motion we will closely follow their derivation, skipping
the details and showing only the main results. The current problem
for the L-junction differs from that of the T-junction microchannel
configuration of Yossifon et Al. 2006 \cite{yossifon_electro-osmotic_2006}
only in that the mid-narrow channel symmetry line is replaced by a
wall (i.e. we apply an electroosmotic slip velocity instead of a symmetry
condition at $D{}_{\infty}E$ boundary, see Fig. \ref{fig:The-L-junction-configuration.}).
It is noted that for a low-conductivity electrolyte, electrothermal
effects stemming from Joule heating (which scales linearly with the
conductivity) become minimal \cite{tang_joule_2006}. Moreover, for
the current geometry, it may also be shown that any inplane temperature
gradients (which would potentially control the 2D electrothermal forcing)
resulting from such minimal Joule heating, would be negligible as
most of the generated heat dissipate through the bottom slide which
has minimum thermal resistance. The introduction into the problem
of a small but finite wall dielectric constant $\lyxmathsym{\textgreek{e}}_{w}$,
necessitates the simultaneous calculation of both $\phi_{f}$ and
$\phi_{w}$ , the electrostatic potentials within the fluid and wall
domains, respectively. On the microscale of the electric double layer
these potentials are coupled through the boundary conditions imposing
continuity of the potential and specifying the jump in electric displacement
across the true solid-liquid interface \cite{landau_electrodynamics_1984}.
The thin-double-layer approximation was previously extended by Yossifon
et al. 2006 \cite{yossifon_electro-osmotic_2006} to account for the
electric-field leakage through the wall for an arbitrary value of
the dielectric constants ratio $\lyxmathsym{\textgreek{e}}_{w}/\lyxmathsym{\textgreek{e}}_{f}$,
to obtain an appropriate macro scale boundary condition for a symmetric
electrolyte solution relating $\phi_{f}$ just outside the double
layer and $\phi_{w}$ at the surface of the wall 

\begin{equation}
\phi_{w}+\alpha\frac{\partial\phi_{w}}{\partial n}=\phi_{f}+\zeta^{eq}\;\; on\;\; A_{\infty}BC_{\infty}.\label{eq:Robin Cond. on wall-1}
\end{equation}
The dimensionless Robin-type boundary condition (\ref{eq:Robin Cond. on wall-1})
is formulated in terms of the dimensionless electric potentials $\phi_{f}$
, $\phi_{w}$ and the equilibrium zeta-potential $\lyxmathsym{\textgreek{z}}^{eq}$
normalized by $E_{0}h_{1}$, wherein $h_{1}$ is the width of the
narrow channel and $E_{0}$ is the magnitude of the externally applied
electric field within the narrow channel far from the junction. Here
$\frac{\partial}{\partial n}$ denotes the derivative in the direction
normal to the wall pointing into the fluid domain. Appearing in (\ref{eq:Robin Cond. on wall-1})
is the parameter $\alpha=\frac{\epsilon_{w}}{\epsilon_{f}}(\kappa h_{1})^{-1}$,
wherein $\kappa^{-1}$ is the (presumed small) Debye length, which
represents the thickness of the electric double layer. Under this
macro scale boundary condition (1) the dimensionless electric potentials
within the bulk fluid and within the wall both satisfy the Laplace
equation. In addition, the potential within the bulk fluid satisfies
the Neumann-type boundary condition $\frac{\partial\phi_{f}}{\partial n}=0$
applied on the channel walls ($A_{\infty}BC_{\infty}$ and $D_{\infty}EF_{\infty}$
in Fig. \ref{fig:The-L-junction-configuration.}). These boundary
conditions are supplemented by the requirements that$\frac{\partial\phi_{f}}{\partial x}=-1$
within the narrow channel far from the junction ( $x\rightarrow-\infty$)
(see Fig. \ref{fig:The-L-junction-configuration.}) and by conservation
of electric-field flux $\frac{\partial\phi_{f}}{\partial y}=-\frac{h_{1}}{h_{2}}$
far from the junction within the wider channel ($y\rightarrow\infty$),
as well as appropriate 'far field' ($x^{2}+y^{2}\rightarrow\infty$)
decay conditions within the wall domain. 

The problems governing $\phi_{f}$ and $\phi_{w}$ are thus decoupled
and may be solved recursively. Making use of the Schwartz-Christoffel
(e.g. Milne-Thomson \cite{milne-thomson_theoretical_2011}) transformation

\begin{equation}
z_{(t)}=\frac{1}{\pi}\left[\frac{h_{2}}{h_{1}}ln\left(\frac{t+1}{t-1}\right)+2tan^{-1}\left(\frac{h_{1}}{h_{2}}t\right)-\pi\right],\label{eq:SC mapping 1}
\end{equation}
and

\begin{equation}
\omega_{(t)}=\frac{1+\left(h_{1}/h_{2}\right)^{2}}{1-t^{2}}.\label{eq:SC mapping 2}
\end{equation}
the physical complex $z=x+iy$ plane is mapped onto the upper half
of the $\omega=\xi+i\eta$ plane, to obtain for a 2D potential of
a point charge located at $\omega=1$

\begin{equation}
\phi_{f}=\frac{1}{\pi}Re\left\{ ln\left(\omega-1\right)\right\} ,\label{eq:Point charge potential in mapped plane}
\end{equation}
wherein $'Re\left\{ \right\} '$ denotes the real part. Expanding
(\ref{eq:SC mapping 1},\ref{eq:SC mapping 2}) and (\ref{eq:Point charge potential in mapped plane})
for $\left|\omega\right|\ll1$ (corresponding to $\left|t\right|\gg\frac{h_{2}}{h_{1}}$
) one obtains the following potential function:

\begin{equation}
\phi_{f}\backsim\frac{1}{\pi}\left(1+\left(\frac{h_{1}}{h_{2}}\right)^{2}\right)^{\unitfrac{1}{3}}Re\left\{ \left(-\frac{3\pi}{2}z\right)^{\unitfrac{2}{3}}\right\} ,\label{eq:Linear symmetric fluid potential}
\end{equation}
which holds in the vicinity of the corner $z=0$. Once $\phi_{f}$
is known, one can calculate $\phi_{w}$ to satisfy Laplace's equation
within the wall domain and (\ref{eq:Robin Cond. on wall-1}) from
which we find the total zeta potential to be
\begin{equation}
\zeta=\phi_{w}-\phi_{f}=\zeta^{eq}+\zeta^{i}\;\; on\;\; A_{\infty}BC_{\infty},
\end{equation}
wherein $\zeta^{i}=-\alpha\frac{\partial\phi_{w}}{\partial n}$ denotes
the induced zeta potential. On the channel walls ($D_{\infty}EF_{\infty}$)
opposite from those forming the corner ($A_{\infty}BC_{\infty}$)
the latter effect is negligible and it is assumed that $\zeta=\zeta_{w}^{eq}$.

The fluid velocity $\mathbf{v}$ and pressure $p$ are normalized
by $\varepsilon_{0}\varepsilon_{f}E_{0}^{2}\nicefrac{h_{1}}{\eta}$
and $\varepsilon_{0}\varepsilon_{f}E_{0}^{2}$, respectively, where
$\eta$ denotes the dynamic viscosity of the electrolyte solution
and $\varepsilon_{0}$ is the permittivity of vacuum. The quasi-steady,
small-Reynolds number flow of the bulk electro-neutral fluid is governed
by the continuity equation $\nabla\cdot\mathbf{v}=0$ and the Stokes
equation $\nabla p=\nabla^{2}\mathbf{v}$. On the channel walls ($A_{\infty}BC_{\infty}$
and $D_{\infty}EF_{\infty}$ in Fig. \ref{fig:The-L-junction-configuration.})
$\mathbf{v}$ satisfies the vanishing of the fluid velocity normal
to the wall and the Helmholtz-Smoluchowski slip velocity \cite{lyklema_fundamentals_1995}condition 

\begin{equation}
\mathbf{v}_{\mathbf{\parallel}}=-\zeta\mathbf{E}_{\mathbf{\parallel}}=-\zeta_{w}^{eq}\mathbf{E}_{\mathbf{\parallel}}-\zeta^{i}\mathbf{E}_{\mathbf{\parallel}},\label{eq:zeta=00003Deq+icd}
\end{equation}
where the subscript $'\Vert'$ denotes the vector component tangent
to the wall. Far upstream and downstream of the corner the pressure
and fluid-velocity vector become uniform across the channel. We therefore
choose to present the velocity field as the sum of the linear electro-osmotic
flow (EOF) and induced contributions

\begin{equation}
\mathbf{v}=\zeta_{w}^{eq}\mathbf{\nabla}\phi_{f}+\mathbf{v_{\mathit{ICEO}}}=\mathbf{\mathbf{v}}_{\mathbf{\mathit{EOF}}}+\mathbf{v_{\mathit{ICEO}}}.\label{eq:V eq+icd}
\end{equation}
The first term on the right-hand side of (\ref{eq:V eq+icd}) is an
irrotational field, and based on the local approximation (\ref{eq:Linear symmetric fluid potential})
may be written as

\begin{equation}
\begin{array}{c}
\mathbf{\mathbf{v}}_{\mathbf{\mathit{EOF}}}=\zeta_{w}^{eq}\mathbf{\nabla}\phi_{f}\\
=\zeta_{w}^{eq}\left(1+\left(\frac{h_{1}}{h_{2}}\right)^{2}\right)^{\unitfrac{1}{3}}\left(\frac{3}{2}\pi r\right)^{-\unitfrac{1}{3}}\left[cos\left(\frac{2}{3}\left(\theta+\pi\right)\right)\hat{\mathbf{r}}-sin\left(\frac{2}{3}\left(\theta+\pi\right)\right)\hat{\mathbf{\mathbf{\mathbf{\boldsymbol{\theta}}}}}\right].
\end{array}\label{eq:linear velocity}
\end{equation}
The induced part of the velocity $\mathbf{v}_{ICEO}$ is derivable
from the stream function $\psi$ satisfying the bi-harmonic equation
$\nabla^{4}\psi=0$ together with the requirements that $\psi=const.$
on each segment of the boundary of the fluid domain ($A_{\infty}BC_{\infty}$
and $D_{\infty}EF_{\infty}$ in Fig. \ref{fig:The-L-junction-configuration.})
and the slip-velocity condition resulting from the second term of
(\ref{eq:zeta=00003Deq+icd}). To obtain an approximate 'local' expression
for $\psi$ in the vicinity of the corner, the plane polar ($r,\theta$)
coordinates were employed such that the origin lies at the corner
and $\theta=0$ corresponds to the bisector of the fluid domain (eq.
3.8' in \cite{yossifon_electro-osmotic_2006}), from which one obtains

\begin{equation}
\begin{array}{c}
\mathbf{v_{\mathit{ICEO}}}=\alpha\left(\frac{3}{2}\right)^{\unitfrac{3}{2}}\left\{ \frac{2}{3\pi}\left[1+\left(\frac{h_{1}}{h_{2}}\right)^{2}\right]\right\} ^{\unitfrac{2}{3}}\\
\cdot\frac{r^{-\unitfrac{2}{3}}}{3}\left[\left(cos\left(\frac{1}{3}\theta\right)+5cos\left(\frac{5}{3}\theta\right)\right)\hat{\mathbf{r}}-\left(sin\left(\frac{1}{3}\theta\right)+sin\left(\frac{5}{3}\theta\right)\right)\hat{\mathbf{\mathbf{\mathbf{\boldsymbol{\theta}}}}}\right].
\end{array}\label{eq:induced velocity}
\end{equation}

\subsection{Particle equation of motion}

In the limit of low particle concentration, particle-particle interactions
may be ignored and the governing kinetic equations for immersed particles
include only terms related to the hydrodynamics of the fluid and externally
applied forces acting on individual particles. The former includes
the ICEO flow and linear EOF, the latter comprised of the DEP and
linear electrophoretic (EP) contribution. For a spherical neutrally
buoyant particle located at $x_{p}\left(t\right)$ and moving with
velocity $\mathbf{v}_{\mathbf{p}}\left(t\right)$ within a fluid flow
$\mathbf{u}$, the dimensionless equation of motion has the form \cite{babiano_dynamics_2000}

\begin{equation}
\begin{array}{c}
St\frac{d\mathbf{v_{p}}}{dt}=St\frac{\rho_{f}}{\rho_{p}}\frac{D\mathbf{u}}{Dt}-St\frac{\rho_{f}}{2\rho_{p}}\left[\frac{d\mathbf{v_{p}}}{dt}-\frac{D}{Dt}\left(\mathbf{u}+\frac{3Fa}{5}\nabla^{2}\mathbf{u}\right)\right]\\
-\left(\mathbf{v_{p}-u-}Fa\nabla^{2}\mathbf{u}\right)-Ba\int_{0}^{t}\left[\frac{1}{\sqrt{t-\tau}}\frac{d}{d\tau}\left(\mathbf{v_{p}-u-}Fa\nabla^{2}\mathbf{u}\right)\right]d\tau+\mathbf{U_{Force}},
\end{array}
\end{equation}
where $\mathbf{U_{Force}}$ is a general term for additional forces
that may act on the particle, e.g. DEP and EP forces in our case.
The typical order of the particle radius $a\backsim O\left(10^{-6}m\right)$,
the channel length scale $L\backsim O\left(10^{-4}m\right)$, velocity
$U_{0}\backsim O\left(10^{-3}\nicefrac{m}{s}\right)$, particle and
fluid densities $\rho_{f},\rho_{p}\backsim O\left(10^{3}\nicefrac{kg}{m^{3}}\right)$
and the fluid dynamic viscosity $\mu\backsim O\left(10^{-3}Pa\cdot s\right)$.
These result in negligible Stokes $\left(St=\frac{2a^{2}\rho_{p}U_{0}}{9\mu L}\thickapprox2\cdot10^{-6}\right)$
, Faxen $\left(Fa=\frac{a^{2}}{6L^{2}}\thickapprox1.7\cdot10^{-5}\right)$
, and Basset $\left(Ba=a\sqrt{\frac{\rho_{f}U_{0}}{\pi\mu L}}\thickapprox1.7\cdot10^{-3}\right)$
dimensionless numbers. Thus, the particle equation of motion can be
reduced to

\begin{equation}
\mathbf{v}_{p}=\mathbf{\mathbf{v}_{\mathit{LINEAR}}}+\mathbf{v_{\mathit{ICEO}}}+\mathbf{\mathbf{v}_{\mathit{DEP}}},\label{eq:velocity components-1}
\end{equation}
wherein the linear component comprises of both EOF and EP terms
\begin{equation}
\mathbf{\mathbf{v}_{\mathit{LINEAR}}}=\mathbf{\mathbf{\mathbf{v}_{\mathit{EOF}}}+\mathbf{\mathbf{v}_{\mathit{EP}}}}=\left(\zeta_{w}^{eq}-\zeta_{p}^{eq}\right)\mathbf{\nabla}\phi_{f}=\zeta^{eq}\mathbf{\nabla}\phi_{f},\label{eq:Linear velocity components}
\end{equation}
with the subscripts $w$ and $p$ standing for the microchannel wall
and particle, respectively. Since both the microchannel wall and particle
surface are negatively charged the EOF and EP counteracts each other
with the former being dominant (see section IV.E and Fig \ref{fig:particle-aggregation-plots}).
To account for particle trapping a non-divergence free attractive
force term must be added \cite{liu_dynamic_2010} which, as was recently
shown for the same tracer particles used in the current study \cite{rozitsky_quantifying_2013,green_dynamical_2013},
is provided by the short-range positive DEP attractive force existing
under DC field conditions. The dimensional (symbolized by tilda) dielectrophoretic
particle velocity contribution is 

\begin{equation}
\mathbf{\tilde{v}}_{\mathbf{\mathit{DEP}}}=\frac{\mathbf{\mathbf{\tilde{F}}}_{\mathbf{\mathit{DEP}}}}{6\pi\eta a}=\frac{\varepsilon_{0}\varepsilon_{f}a^{2}f_{CM}\tilde{\nabla}\left|\mathbf{\tilde{E}}\right|^{2}}{3\eta},\label{eq:DEP dimensional velocity}
\end{equation}
wherein $a$ is the particle radius and $f_{CM}$ is the Clausius-Mossotti
factor \cite{jones_electromechanics_1995}. Non-dimensionalization
yields 

\begin{equation}
\mathbf{v}_{\mathbf{\mathit{DEP}}}=\frac{1}{3}f_{CM}\left(\frac{a}{h_{1}}\right)^{2}\nabla\left|\mathbf{E}\right|^{2},\label{eq:DEP nondimensionalization}
\end{equation}
where from the local approximation of the bulk fluid potential (\ref{eq:Linear symmetric fluid potential})

\begin{equation}
\mathbf{v}_{\mathbf{\mathit{DEP}}}=-f_{CM}\left(\frac{a}{h_{1}}\right)^{2}\frac{\pi}{3}\left[1+\left(\frac{h_{1}}{h_{2}}\right)^{2}\right]^{\unitfrac{2}{3}}\left(\frac{3\pi r}{2}\right)^{\unitfrac{-5}{3}}\hat{\mathbf{r}.}\label{eq:DEP  velocity}
\end{equation}
Thus, the total particle velocity (\ref{eq:velocity components-1})
can be rewritten in terms of the \textit{relative }contributions of
the above linear (\ref{eq:linear velocity},\ref{eq:Linear velocity components}),
induced (\ref{eq:induced velocity}) and dielectrophoretic(\ref{eq:DEP  velocity})
velocities, 

\begin{equation}
\frac{\mathbf{v}_{p}}{\zeta^{eq}}=\mathbf{\mathbf{\nabla}}\phi_{f}+\lambda_{ICEO}\frac{\mathbf{v}_{ICEO}}{\alpha}+\lambda_{DEP}\frac{\mathbf{v}_{DEP}}{f_{CM}\left(\nicefrac{a}{h_{1}}\right)^{2}},\label{eq:velocity components lambdas}
\end{equation}
where $\lambda_{ICEO}=\left|\nicefrac{\alpha}{\zeta^{eq}}\right|$
represents the relative importance of the ``induced'' and ``linear''
respective parts of $\mathbf{v}_{p}$, while $\lambda_{DEP}=\left|\left(\frac{a}{h_{1}}\right)^{2}\nicefrac{f_{CM}}{\zeta^{eq}}\right|$
represents the relative importance of its ``dielectrophoretic''
and ``linear'' respective parts.

\section{Results and Discussion}

\subsection{Experimental}

\begin{figure}
\begin{centering}
\includegraphics[width=0.95\columnwidth]{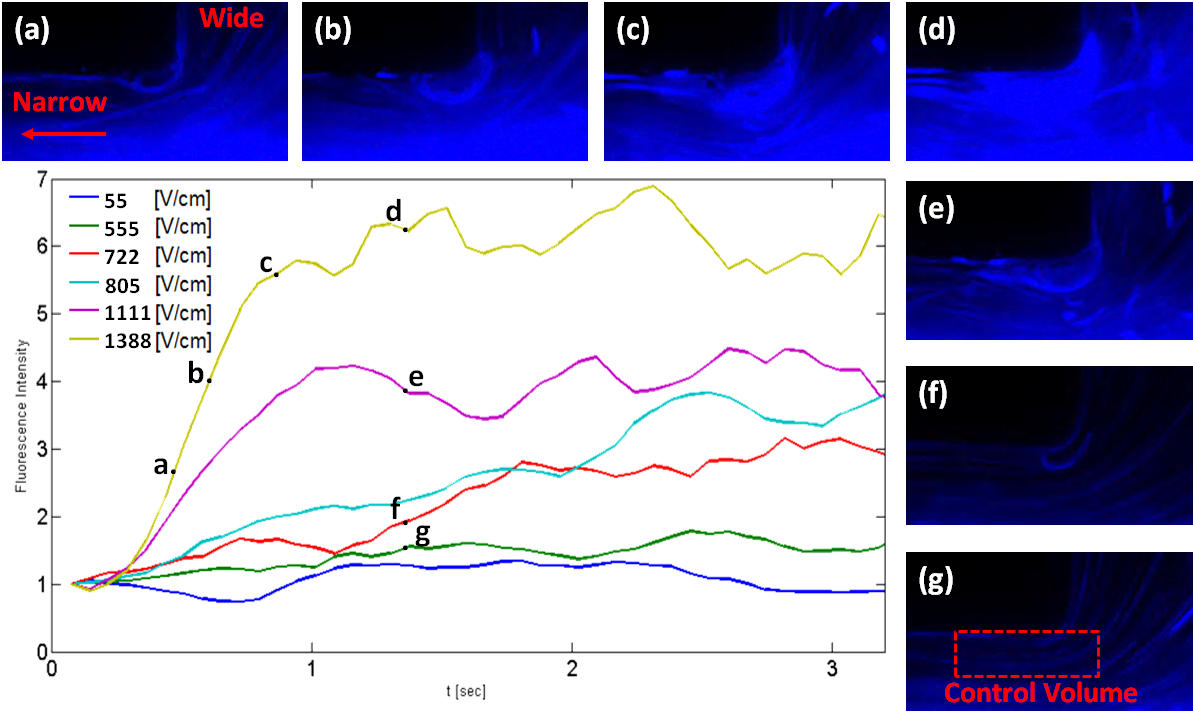}
\par\end{centering}

\caption{\label{fig:particle-aggregation-plots}Time evolution of the total
fluorescent intensity (normalized by that corresponding to t=0 {[}s{]})
within the control volume (depicted in inset g) obtained using Confocal
microscopy for blue 1{[}\textgreek{m}m{]} fluorescent particles with
$\frac{h_{2}}{h_{1}}=5$. Insets a-d are confocal images taken at
varying time under a constant applied field of 1388 {[}V/cm{]}, while
insets d-g correspond to a constant time of \textasciitilde{}1.3{[}s{]}
at various applied fields. The arrow in inset a of the figure indicate
the direction of the electric field (i.e. $E_{0}<0$) and the net
particle motion away from the corner.}
 
\end{figure}

The time evolution of the particle trapping is quantified in Fig.
\ref{fig:particle-aggregation-plots}
 by determining the overall fluorescent intensity within a constant
control volume for various applied electric fields ($E_{0}<0$) and
particle sizes. The particle net motion away from the corner is in
the direction of the electric field indicating the dominance of EOF
over EP. Images were obtained using the confocal microscope system.
It is clearly demonstrated, based on the slope of the curves in Fig.
\ref{fig:particle-aggregation-plots}, that the accumulation becomes
more rapid as the voltage increases. Also, it is seen that for low
enough fields (e.g. under $\sim$555 {[}V/cm{]}) the accumulation
is almost negligible. This observation is in qualitative agreement
with (\ref{eq:DEP nondimensionalization}) where it can be seen that
the DEP trapping force scales non-linearly with the electric field.
The insets a-g in Fig. \ref{fig:particle-aggregation-plots} are confocal
images taken at discrete times and various voltages to illustrate
the particle accumulation at the corner vicinity assisted by an ICEO
vortex downstream of the corner. Insets a-d, corresponding to a constant
applied field of 1388 {[}V/cm{]}, depict the time accumulation of
the particles until a saturation like behavior occurs (at \textasciitilde{}1.3
{[}s{]}) which may correspond to a finite capacity of trapped particles
(see Movie\#1 in \cite{_supplementary_2013}). Insets d-g are images
taken at time of \textasciitilde{}1.3 {[}s{]}, which depict the downstream
vortex patterns and particle accumulation for various applied fields.
It is clear from Fig. \ref{fig:particle-aggregation-plots} that saturation
is reached at longer times with decreased applied fields as can be
expected from the corresponding observed decrease in vortex intensities
along with particle trapping (insets d-g).

\begin{figure}
\begin{centering}
\includegraphics[width=0.95\columnwidth]{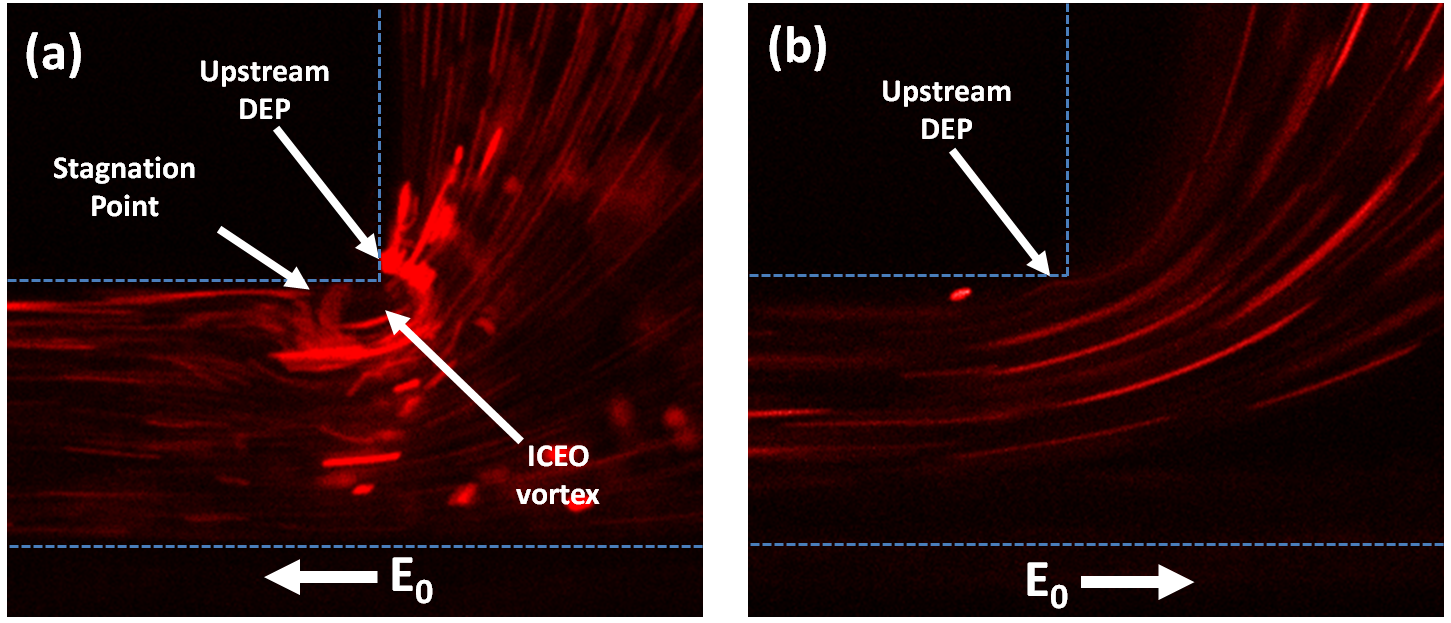}
\par\end{centering}

\caption{\label{fig:Asymmetry Experimental}Particle dynamics of fluorescent
red 1{[}\textgreek{m}m{]} particles upon reversal of the electric
field polarity (bold arrow). $\frac{h_{2}}{h_{1}}=5$ , externally
applied field $\left|E_{0}\right|$ is 833 {[}V/cm{]}.}
\end{figure}

\begin{figure}
\begin{centering}
\includegraphics[width=0.95\columnwidth]{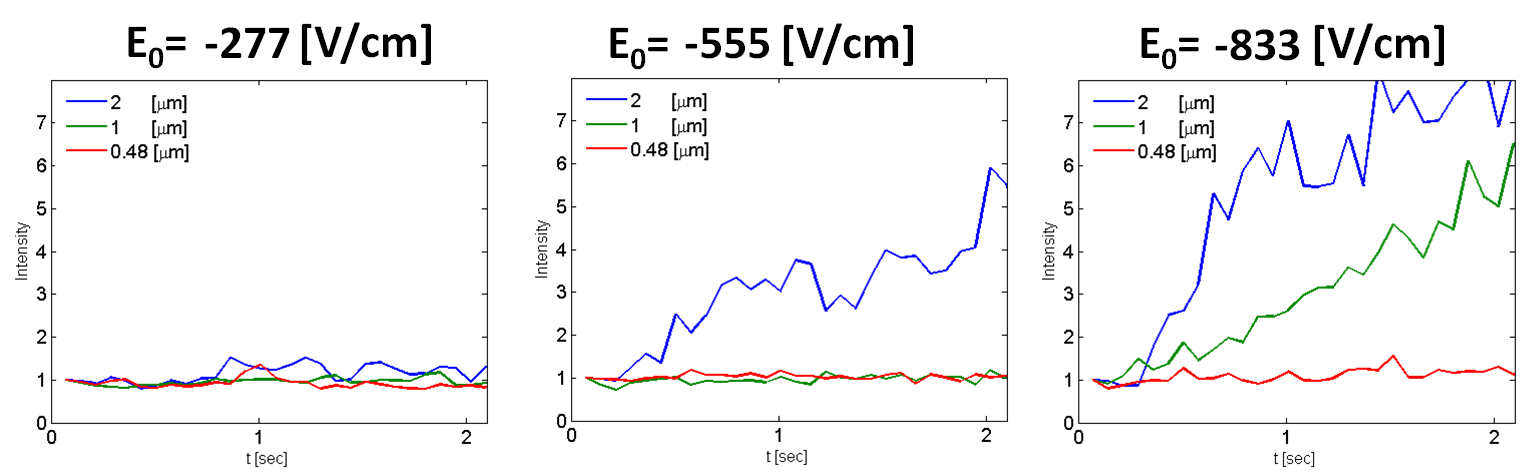}
\par\end{centering}

\caption{\label{fig:Particle size effects}The effect of the particle size
on time evolution of the total fluorescent intensity (normalized by
that corresponding to t=0 {[}s{]}) at various applied electric fields.}
\end{figure}

Curiously, upon reversal of the electric field direction, a pronounced
asymmetry of the downstream ICEO vortex together with particle trapping
(Fig. \ref{fig:Asymmetry Experimental})
 are observed. For an electric field directed from the narrow to wide
channel ($E_{0}>0$) no trapping is seen downstream of the corner
and the dark region presumably consists of the downstream vortex (see
Movie\#2 in \cite{_supplementary_2013}). As will be shown theoretically
below, for $E_{0}>0$, DEP is no longer effective at trapping particles
downstream of the corner. However, upstream of the corner, DEP trapping
is observed irrespective of field direction. That the trapping increases
with the particle size (Fig. \ref{fig:Particle size effects})
 is yet another indication that the mechanism is DEP controlled as
the force scales linearly with the particle volume (\ref{eq:DEP dimensional velocity})
(see Movie \#3 in \cite{_supplementary_2013}). In the above figures
the electric fields within the narrow channels were calculated based
on the geometry of the two channels and the applied voltage according
to a simplified Ohmic model $E_{0}=V/\left(L\left(1+h_{1}/h_{2}\right)\right)$.
$L$ is the length of the narrow and wide channels (i.e. 15mm) and
$V$ is the applied potential difference across the reservoirs.

\subsection{Theoretical model}

\begin{figure}
\begin{centering}
\includegraphics[width=0.95\columnwidth]{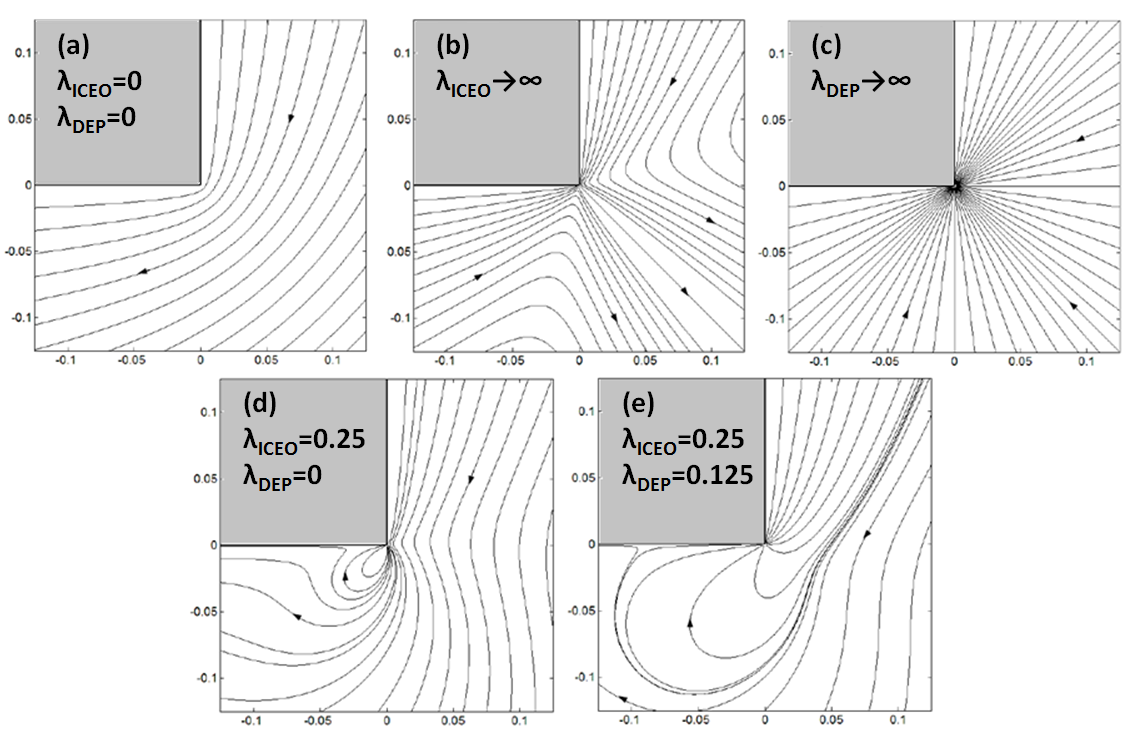}
\par\end{centering}

\caption{\label{fig:gilads-fields} Analytically derivable local flow streamlines
(insets (a), (b) and (d)) and particle pathline patterns (insets (c)
and (e)) in the vicinity of the corner for $\frac{h_{2}}{h_{1}}=5$,
$E_{0}<0$, and the indicated values of $\lambda_{ICEO}$ and $\lambda_{DEP}$.}
\end{figure}

We examine the evolution of the topology of the particle pathlines
in the vicinity of the channel corner for varying ratios $\lambda_{ICEO}=\left|\nicefrac{\alpha}{\zeta^{eq}}\right|$
and $\lambda_{DEP}=\left|\left(\frac{a}{h_{1}}\right)^{2}\nicefrac{f_{CM}}{\zeta^{eq}}\right|$.
Alternatively, by the assumed scaling of $\zeta^{eq}$, the following
also corresponds to the development of the flow and particle dynamics
with increasing intensity of the external field. Fig. \ref{fig:gilads-fields}
presents the local particle pathline pattern in the vicinity of the
corner (i.e., $r\ll1$) for $\frac{h_{2}}{h_{1}}=5$ at the indicated
values of $\lambda_{ICEO}$ and $\lambda_{DEP}$. Qualitatively similar
figures are obtained when selecting other values of $\frac{h_{2}}{h_{1}}$.
Inset (a) depicts the irrotational antisymmetric velocity field (see
the first term in (\ref{eq:velocity components lambdas}) ) corresponding
to the linear part of $\mathbf{v}_{p}$ ($\lambda_{ICEO}=0,\,\lambda_{DEP}=0$),
which is derivable from $\phi_{f}$ (\ref{eq:Linear symmetric fluid potential})
and represents the superposition of EOF and opposing EP. The induced
part ($\lambda_{ICEO}\rightarrow\infty$) of the electro-osmotic flow,
$\mathbf{v}_{\mathbf{\mathit{ICEO}}}$, is presented in part (b).
The symmetric field obtained from (\ref{eq:induced velocity}) demonstrates
the formation of a jet by the convergent flow at the corner. Inset
(d) of the same figure shows the flow resulting from the superposition
of both the linear and induced flow components for $\lambda_{ICEO}=0.25$.
As $r\rightarrow0$ the induced velocity ($\varpropto r^{-\unitfrac{2}{3}}$)
prevails over the linear ($\varpropto r^{-\unitfrac{1}{3}}$) part.
Accordingly, it produces a reverse flow along the wall immediately
downstream of the corner. This is expected to create a domain of closed
streamlines as indeed appears in Fig. \ref{fig:gilads-fields}(d)
. The extent of this domain is diminishing with decreasing $\lambda_{ICEO}$,
or equivalently with decreasing electric field intensity, which is
in qualitative agreement with the experimental results shown in insets
(d-g) of Fig. \ref{fig:particle-aggregation-plots}. Upstream of the
corner, the superposed velocities accumulate and for this reason no
reverse flow appears at the wall. As shown in the sequel, reverse
flow and closed-streamline domains may nevertheless occur upstream
of the corner (Fig. \ref{fig:COMSOL results}(a,c) and Yossifon et
al. 2006 \cite{yossifon_electro-osmotic_2006}) for sufficiently large
values of $\lambda_{ICEO}$. However, unlike the vortex occurring
downstream of the corner, this reverse flow occurs near the opposite
channel walls ($D_{\infty}EF_{\infty}$) and is therefore not a \textquotedblleft{}local\textquotedblright{}
phenomenon. Inset (c) represents the motion of the particles dominated
solely by DEP ($\lambda_{DEP}\rightarrow\infty$ ) whereby the colloids
are attracted radially towards the corner acting as a sink. Finally,
the particle pathlines resulting from the superposition of the dielectrophoretic
contribution ($\lambda_{DEP}=0.125$) to the ICEO fluid flow ($\lambda_{ICEO}=0.25$)
are shown in inset (e). Here, as $r\rightarrow0$ it is the DEP velocity
that prevails ($\varpropto r^{-\unitfrac{5}{3}}$) over both the induced
($\varpropto r^{-\unitfrac{2}{3}}$) and linear ($\varpropto r^{-\unitfrac{1}{3}}$)
velocity components. This manifests itself in attracting pathlines
which are downstream towards the corner where particles are trapped.
Additionally, the DEP force \textquotedblleft{}breaks\textquotedblright{}
the closed streamlines obtained for the ICEO flow (inset (d)) at the
downstream side of the corner. This results in significant enhancement
to the trapping rate due to particles arriving from the upstream side.
This prediction is in agreement with the experimental results shown
in Figs. \ref{fig:particle-aggregation-plots},\ref{fig:Asymmetry Experimental}
which illustrate the DEP trapping of the particles both at the upstream
side of the corner to which particles pathlines are attracted and
at its downstream side where rapid accumulation of particles is assisted
by the ICEO vortex. The linear relationship between DEP force and
particle volume (\ref{eq:DEP  velocity}) is clearly visible in the
experimental results (Fig. \ref{fig:Particle size effects}) wherein
the rate of accumulation increases with increasing particle size.

\subsection{Asymmetry of the problem}

An intriguing feature of the experimental results is the distinct
difference in particle dynamics and their eventual accumulation upon
reversal of the electric field polarity (see Fig. \ref{fig:Asymmetry Experimental}).
The theoretical predictions based on the approximate 'local' expression
of the potential (\ref{eq:Linear symmetric fluid potential}); Fig.
\ref{fig:gilads-fields}), which is symmetric with respect to $\theta$,
cannot account for this effect which stems from the geometrical asymmetry
of the L-junction microchannels width $\left(\frac{h_{2}}{h_{1}}\neq1\right)$.
In contrast, eqs.(\ref{eq:SC mapping 1}-\ref{eq:Point charge potential in mapped plane})
are the 'global' solution of the potential from which the linear asymmetric
particle velocity field can be directly extracted to obtain

\begin{subequations}
\begin{equation}
\mathbf{v}_{LINEAR}=\zeta^{eq}\mathbf{\mathbf{\nabla}}\phi_{f}=-\zeta^{eq}\frac{1}{\pi}Re\left\{ \frac{1}{\omega-1}\frac{d\omega}{dt}\frac{dt}{dz}\nabla z\right\} ,\label{eq:asymmetric linear - derivatives}
\end{equation}
which from equations (\ref{eq:SC mapping 1},\ref{eq:SC mapping 2})
yields

\begin{equation}
\mathbf{v}_{LINEAR}=\zeta^{eq}\frac{h_{1}}{h_{2}}\left[Re\left\{ it\right\} \hat{\mathbf{x}}-Re\left\{ t\right\} \hat{\mathbf{y}}\right].\label{eq:asymmetric linear velocity}
\end{equation}

\end{subequations}

\begin{figure}
\begin{centering}
\includegraphics[width=0.95\columnwidth]{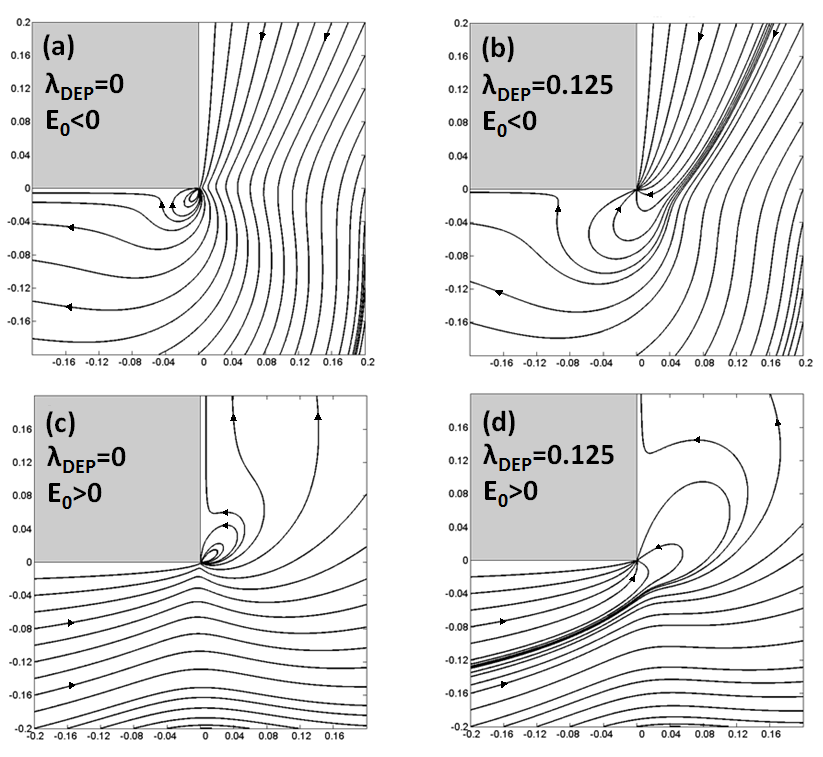}
\par\end{centering}

\caption{\label{fig:asymmetric  field}Particle pathlines upon reversal of
the applied electric field polarity following an asymmetric linear
field with symmetric DEP and ICEO fields, in the vicinity of the corner
for $\frac{h_{2}}{h_{1}}=5$ and $\lambda_{ICEO}=0.25$. The narrow
channel is on the bottom-left while the wide channel is on the top-right
(see Fig. \ref{fig:The-L-junction-configuration.}).}
\end{figure}

We combine this asymmetric linear field (\ref{eq:asymmetric linear velocity})
with the induced (\ref{eq:induced velocity}) and DEP (\ref{eq:DEP  velocity})
velocities in their symmetric form as they decay more rapidly then
the linear part ($\varpropto r^{-\unitfrac{5}{3}}$ and $\varpropto r^{-\unitfrac{2}{3}}$,
as opposed to $\varpropto r^{-\unitfrac{1}{3}}$ respectively). These
combined fields were then calculated in the vicinity of the corner
for both polarities of the electric field and are illustrated in Fig.
\ref{fig:asymmetric  field}
. From Fig. \ref{fig:asymmetric  field} it is seen that qualitatively,
the DEP trapping mechanism is similar to that described in the symmetric
case (Fig. \ref{fig:gilads-fields}(e)), except that when the field
is directed from the narrow to wide channels (Fig. \ref{fig:asymmetric  field}(b))
the resulting downstream vortex is larger and so is the region of
particle pathlines downstream of the corner that terminate at the
corner tip (Fig. \ref{fig:asymmetric  field}(d)). It is hard to explain
based only on this why there is such a distinct difference in the
particle entrapment in Fig. \ref{fig:Asymmetry Experimental}. However,
it is clear that the enlargement of the downstream vortex size in
the wider channel compared to that in the opposite direction, due
to the asymmetric linear velocity field, occurs concurrently with
the decrease of the DEP force due to the same asymmetry of the electric
field. With further increase of the electric field, a 'global' solution
accounting for the finite dimensions of the channel's width would
further emphasize this asymmetry, as will be shown in Fig. \ref{fig:COMSOL results},
where the size of the vortex in the narrow channel is shown to quickly
approach its maximum size which is dictated by the channel width.

\subsection{Numerical results}

\begin{figure}
\begin{centering}
\includegraphics[width=0.95\columnwidth]{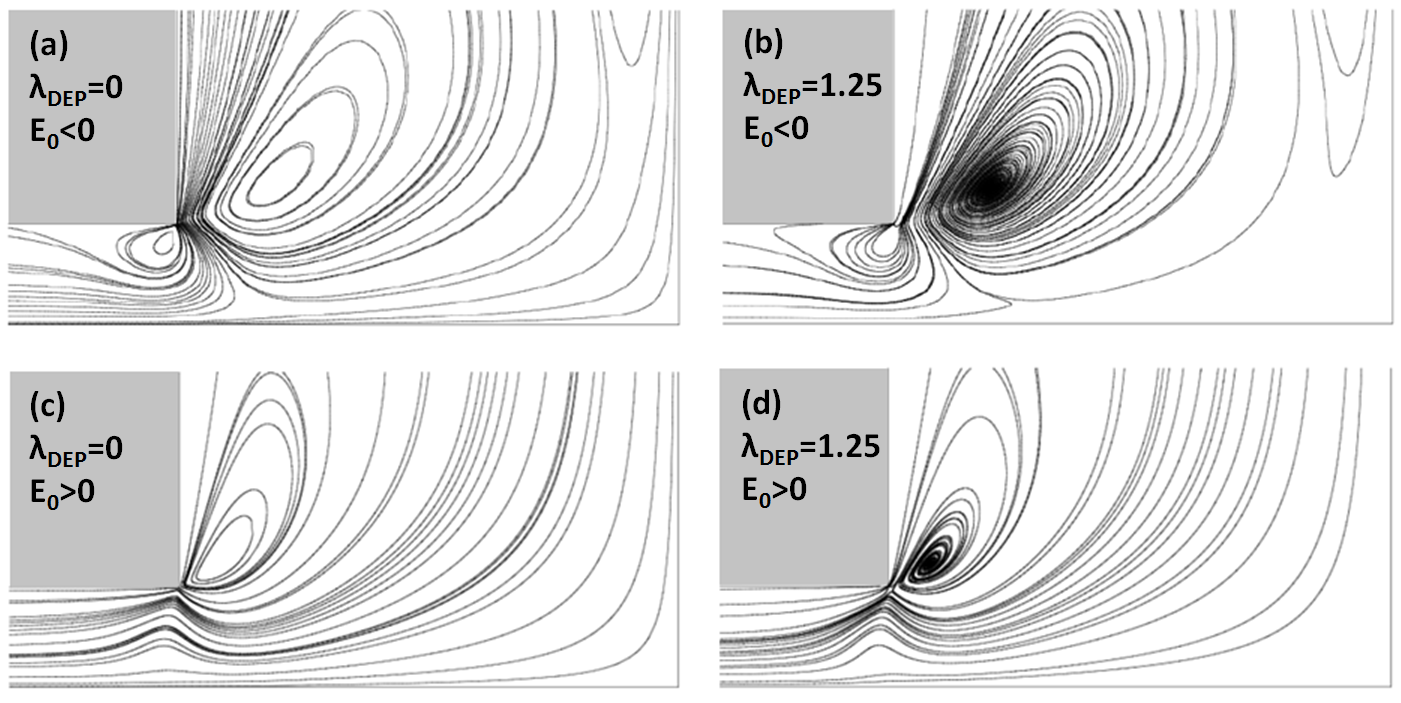}
\par\end{centering}

\caption{\label{fig:COMSOL results} \textquotedblleft{}Global\textquotedblright{}
numerical solution of the particle pathlines patterns upon reversal
of the applied electric field polarity for $\frac{h_{2}}{h_{1}}=5$
and $\lambda_{ICEO}=2.5$,}
\end{figure}

Using commercial code (COMSOL) the decoupled electrostatic (Laplace
equations within the bulk fluid and wall domains along with (\ref{eq:Robin Cond. on wall-1}))
and hydrodynamic (Stokes and continuity equations along with (\ref{eq:zeta=00003Deq+icd}))
equations were solved within the full microchannel domain) rather
than only at the very vicinity of the corner as in the analytically
derived eqs. (\ref{eq:linear velocity},\ref{eq:induced velocity},\ref{eq:DEP  velocity}).
Upon this solution, the particle DEP velocity component (\ref{eq:DEP nondimensionalization})
was added in order to obtain the particle streamlines. Fig. \ref{fig:COMSOL results}
 provides a \textquotedblleft{}global\textquotedblright{} description
of the streamline pattern for $\lambda_{ICEO}=2.5$ for two opposite
electric field polarities with/without the DEP contribution. insets
(a) and (c) of the figure show the vortex formation downstream of
the corner due to the combined induced and linear flows under reverse
field polarity. While the induced ejection intensity is the same in
both cases, the linear flows (and the electric fields) roughly differ
by a factor of $\tfrac{h_{2}}{h_{1}}$, which corresponds to the ratio
of the average velocities within the narrow and wide channels away
from the corner, affecting the size of the formed downstream vortex.

Insets (b) and (d) of the figure show the particle pathlines corresponding
to the flow fields in insets (a) and (c), respectively, with the addition
of the DEP force. That DEP force attracts particles towards the corner
is clearly seen both in the upstream side where streamlines are terminated
at the corner and the downstream side where the streamlines (insets
(b) and (d)) are also terminated at the corner after converging along
the corner wall. These simulation results are in good agreement with
both the theoretical (Fig. \ref{fig:gilads-fields}, Fig. \ref{fig:asymmetric  field})
results, as well as experimental observations (Fig. \ref{fig:Asymmetry Experimental}).
Interestingly, upon reversal of the field polarity (inset (d)) the
particle streamlines look very different to those in inset (b). This
stands in contrast to the analytically predicted asymmetric plot (Fig.
\ref{fig:asymmetric  field}), where although the size of the downstream
vortex was different in correspondence with the channel width, the
same qualitative behavior of particle trapping was predicted. This
discrepancy can be explained by the fact that the results shown in
Fig. \ref{fig:COMSOL results} were calculated for relative high applied
fields wherein the 'global' solution was strongly influenced by the
``far'' boundaries. In contrast, despite the global linear contribution,
the analytical solution remains local (Fig. \ref{fig:asymmetric  field})
due to the DEP and ICEO local approximations. This was confirmed by
numerical simulations at low enough voltages (Fig. S1 in \cite{_supplementary_2013})
which approximate the local solution of Fig. \ref{fig:asymmetric  field}.

With the inclusion of the DEP force, the vortex center points (both
upstream and downstream) become unstable. In insets (a),(c) the vortices
have closed stream/pathlines, as opposed to (b),(d) where particles
are moving in a spiral path into/away from the vortex center. This
break in the stability of the fixed points is due to the attracting
p-DEP force at the corner which leads to their conformation change
from closed streamlines to unstable spirals that eject particles from
the vortex center. The above arguments support the experimental findings
(Fig. \ref{fig:Asymmetry Experimental}) where particles were not
observed to be trapped upstream of the corner in the case of the field
pointing from the narrow to wide channels (i.e. $E_{0}>0$), in contrast
to the immense trapping occurring in the opposite direction.

Inset (c) here resembles Fig. 4(c) in \cite{yossifon_electro-osmotic_2006}
in the sense that an upstream vortex may develop, except that the
same flow pattern occurs at yet larger $\lambda_{ICEO}$ values (or
equivalently larger electric fields). The difference in the $D_{\infty}E$
(bottom) boundary condition (H-S slip condition at the wall instead
of a symmetry line) delays the flow reversal within the narrow channel
upstream of the corner, resulting from the continuity of the mass
flow rate, which is necessary for obtaining the downstream vortex.

\subsection{Scaling arguments regarding the different competing mechanisms}

Here we use the following quantities: $\epsilon_{0}=8.85\cdot10^{-12}\unitfrac{F}{m}$
is the vacuum permittivity; $\epsilon_{f}=80$ and $\epsilon_{w}=3$
are the relative dielectric constants of the fluid and microchannel
(PDMS), respectively; $R=8.314\unitfrac{J}{\left(K\cdot mol\right)}$
is the universal gas constant; $T=300^{\circ}K$ is the absolute temperature;
$F=9.648\cdot10^{4}\unitfrac{C}{mol}$ is the Faraday constant; $c_{0}=10^{-5}M$
is the bulk ionic concentration; $h_{1}=80\cdot10^{-6}m$ is the narrow
channel width; $a=0.5\cdot10^{-6}m$ is the particle radius; $\sigma_{p}=133.5\unitfrac{\mu S}{cm}$
(obtained for DI, i.e. $\sigma_{f}=0.05\unitfrac{\mu S}{cm}$\cite{rozitsky_quantifying_2013})
and $\sigma_{f}=2\unitfrac{\mu S}{cm}$ (measured) are the conductivities
of the particle (effective conductivity due to surface conduction)
and medium; the zeta potential of the PDMS can be estimated as\cite{kirby_zeta_2004}
$\tilde{\zeta}_{w}^{eq}=-50mV$ for pH$\thickapprox$5.5 (measured)
while that of polystyrene particles $\widetilde{\zeta}_{p}^{eq}=-34mV$
(measured using Zetasizer for 1\textgreek{m}m particles in $3\cdot10^{-5}M$
solution, $\sigma_{f}=4\unitfrac{\mu S}{cm}$).

Hence, $\kappa^{-1}=\lambda_{D}=\sqrt{\frac{\epsilon_{0}\epsilon_{f}RT}{2F^{2}c_{0}}}\thickapprox97nm$
is the Debye length. Using the following expressions: $\alpha=\frac{\epsilon_{w}\lambda_{D}}{\epsilon_{f}h_{1}}$,
$\zeta^{eq}=\frac{\tilde{\zeta}^{eq}}{E_{0}h_{1}}$ (wherein the tilde
stands for the dimensional zeta-potential) and $f_{CM}=\frac{\sigma_{p}-\sigma_{f}}{\sigma_{p}+2\sigma_{f}}\approx0.96$
then for a value of $E_{0}=1388\unitfrac{V}{cm}$ (corresponding to
the case when the combined effect of all competing mechanisms is most
clearly seen experimentally in Fig.3): 

\begin{subequations}

\begin{equation}
\lambda_{ICEO}=\frac{\alpha}{\zeta^{eq}}=\frac{\alpha}{\zeta_{w}^{eq}-\zeta_{p}^{eq}}=\frac{\epsilon_{w}\lambda_{D}}{\epsilon_{f}\left(\tilde{\zeta_{w}^{eq}}-\tilde{\zeta_{p}^{eq}}\right)}E_{0}\approx0.028
\end{equation}

\begin{equation}
\lambda_{DEP}=\left(\frac{a}{h_{1}}\right)^{2}\frac{f_{CM}}{\zeta^{eq}}=\frac{a^{2}}{h_{1}}\frac{f_{CM}}{\left(\tilde{\zeta_{w}^{eq}}-\tilde{\zeta_{p}^{eq}}\right)}E_{0}\approx0.023
\end{equation}

\end{subequations}

Interestingly, the ratio $\lambda_{ICEO}/\lambda_{DEP}\approx1.21$
is in agreement with the theoretical analysis wherein a ratio of the
$O\left(1\right)$ was used. These values are smaller by factors of
$\thickapprox$9 and $\thickapprox$5 than the theoretical values
of $\lambda_{ICEO}=0.25$ and $\lambda_{DEP}=0.125$ (Fig.6), corresponding
to the case when all competing mechanisms exist simultaneously to
obtain qualitatively the combined effect observed experimentally at
high applied fields ( $E_{0}=1388\unitfrac{V}{cm}$). Such a discrepancy
is not unexpected in the very crude estimates of scaling analysis.
However, one plausible explanation for this could be that while DEP,
ICEP and EP are local effects (the former two are non-negligible only
at the vicinity of the corner while the latter depends only on the
local electric field) the electroosmotic flow must obey the continuity
equation (i.e. mass flux balance) everywhere in the system. Thus,
it is more prone to disturbances, .e.g reservoir-microchannel entrance
effects inducing back pressure which may lower the net EOF flow, resulting,
in better agreement between theory and experiments.

\section{Concluding remarks}

The purpose of the present contribution is to study the colloid dynamics
in conjunction with appearance of ICEO vortices in a micro-channel
junction configuration. Our main experimental results appear in Fig.
\ref{fig:particle-aggregation-plots} where we demonstrate the rapid
particle trapping and accumulation at the vicinity of the corner.The
exact nature of the interparticle forces {[}e.g. Posner 2009\cite{posner_properties_2009}{]}
existing at the corner vicinity when the particle accumulation becomes
significant is beyond the scope of the current study as we are mainly
concerned with the physical mechanism that is responsible, in the
first place, for the migration of particles towards the corner where
they can potentially accumulate. This phenomenon was shown theoretically,
at the single particle level, to be due to the combination of a short-range
non-divergence free DEP trapping force which is assisted by a far-field
electro-convection downstream vortex that feeds the corner with particles
(Fig. \ref{fig:gilads-fields}). The analytical model is a local approximation
of the global solution in the vicinity of the corner. Accordingly,
the resulting expressions involve no indeterminate parameters whose
evaluation would require the use of various ad hoc estimates, in particular
with respect to obtaining a correct description of the dielectrophoretic
contribution to the particle velocity superposed on the ICEO and linear
flow velocities. Finally, we clearly demonstrate that the sharp corner
geometry, in addition to its potential to greatly enhance forced convection
due to the intensified velocity near the channel walls around the
induced vortex, can also be used for rapid trapping of colloids.

\section*{Acknowledgments}

We thank Alicia Boymelgreen for her invaluable inputs. This work was
supported by ISF grant 1078/10. The fabrication of the chip was possible
through the financial and technical support of the Technion RBNI (Russell
Berrie Nanotechnology Institute) and MNFU (Micro Nano Fabrication
Unit).

\end{document}